\newif\ifsubmode
\submodetrue
 
\newif\ifprintfig
\printfigfalse
 
\ifsubmode
  \documentclass{aastex}
  \usepackage{epsfig}
  \received{}
  \accepted{}
  \journalid{}{}
  \articleid{}{}
\else
  \documentclass[preprint]{aastex}
  \usepackage{epsfig}
\fi
 
\shortauthors{B\"oker \ea}
\shorttitle{NICMOS Detectors in Space}

\newcommand{\ea}{et al.}

\newcommand{\K}{\>{\rm K}}
\newcommand{\KPD}{\>{\rm K\,day^{-1}}}
\newcommand{\s}{\>{\rm s}}
\newcommand{\minute}{\>{\rm min}}

\newcommand{\km}{\>{\rm km}}

\newcommand{\mm}{\>{\rm mm}}
\newcommand{\mum}{\>{\mu {\rm m}}}
\newcommand{\eps}{\>{\rm e^-s^{-1}}}

\newcommand{\bdm}{\begin{displaymath}} 
\newcommand{\edm}{\end{displaymath}}
\newcommand{\beq}{\begin{equation}} 
\newcommand{\eeq}{\end{equation}} 
\newcommand{\bit}{\begin{itemize}} 
\newcommand{\eit}{\end{itemize}} 
\newcommand{\ben}{\begin{enumerate}} 
\newcommand{\een}{\end{enumerate}}
\newcommand{\bfi}{\begin{figure}[htb]} 
\newcommand{\bpfi}{\begin{figure}[p]}

\begin{document}

\title{Properties of PACE-I HgCdTe Detectors in Space - the NICMOS
Warm-up Monitoring Program}

\author{T. B\"oker\altaffilmark{1}, J. Bacinski, L. Bergeron, D. Calzetti,
M. Jones, D. Gilmore, S. Holfeltz, B. Monroe, A. Nota,\altaffilmark{1}, 
M. Sosey} 
\affil{Space Telescope Science Institute, 3700 San Martin Drive, 
       Baltimore, MD 21218}
\email{boeker@stsci.edu}

\author{G. Schneider, E. O'Neil, P. Hubbard, A. Ferro, I. Barg, E. Stobie} 
\affil{Steward Observatory, University of Arizona, 933 North Cherry Avenue, 
Tucson, AZ 85721}


\altaffiltext{1}{Affiliated with the Astrophysics Division, Space Science 
Department, European Space Agency.} 

 
\ifsubmode\else
\clearpage\fi
 
 
\ifsubmode\else
\baselineskip=14pt
\fi
 
 
\begin{abstract}
We summarize the results of a monitoring program which was
executed following the cryogen exhaustion of the Near Infrared Camera
and Multi-Object Spectrometer (NICMOS) onboard the Hubble Space Telescope.
During the subsequent warm-up, detector parameters such as 
detective quantum efficiency, dark current, 
bias offsets, and saturation levels have been 
measured over the temperature range $62\K$ to about $100\K$. The measurements
provide a unique database of the characteristics of
PACE-I HgCdTe detector arrays in the space environment. A surprising
result of the analysis is the fact that all three NICMOS detectors
showed an enhanced dark current in the temperature range 
between $77\K$ and $85\K$. However, a subsequent laboratory experiment
designed to replicate the on-orbit warm-up
did not reproduce the anomaly, despite the fact that it employed
a flight-spare detector of the same pedigree. 
The mechanism behind the on-orbit dark current anomaly is therefore believed
to be unique to the space environment.
We discuss possible explanations for these unexpected observational results,
as well as their implications for future NICMOS operations.
\end{abstract}


\keywords{instrumentation:detectors}
          
\clearpage


\section{Introduction}
The Near Infrared Camera and Multi-Object Spectrometer NICMOS 
\citep{tho92,tho98,boe00},
was installed onboard the Hubble Space Telescope (HST) during
the second Servicing Mission in February 1997. NICMOS provides HST 
with infrared imaging and spectroscopic capabilities at wavelengths
between 0.8 and $2.5\mum$. It offers three cameras with different 
image scales and adjacent, but not contiguous fields. Each of the three
cameras is equipped with a NICMOS-3 type, $256\times 256$ pixel
HgCdTe-detector manufactured by Rockwell. 

Shortly after its on-orbit installation, it was discovered that the 
NICMOS dewar suffered from a thermal anomaly that led to a higher 
than expected sublimation rate of the solid nitrogen coolant, and 
thus a shortened lifetime of the instrument. 
After an intensified science program, operation of NICMOS for astronomical 
observations was suspended on Dec. 18, 1998. The cryogen
was depleted on Jan. 4, 1999. Since then, the instrument has been
inactive except for engineering telemetry data
at the ambient temperature of the HST aft shroud ($\approx 280\K$).

Soon after the shortened NICMOS lifetime became apparent,
NASA started investigations of possible means to continue NICMOS
operation. As a result of this process, the NICMOS Cooling System
\citep[NCS,][]{che98} 
will be installed during the next HST servicing mission
late in 2001 in order to maintain an infrared capability
on HST. The NCS is a mechanical cooler which uses a closed-loop
reverse-Brayton cycle to maintain the NICMOS detectors at 
temperatures around $75\K$. The detectors when cooled by the 
NCS will therefore be 
$15-20\K$ warmer than during the solid nitrogen period.

In preparation for NICMOS/NCS operation, a warm-up monitoring 
program was initiated immediately after the NICMOS science program was 
completed. The main goal of this program was to utilize the one-time 
opportunity of the instrument warm-up to monitor the performance
of the NICMOS detectors as their temperatures passed through the range
expected for operations under the NCS. A good understanding of the
temperature dependence of NICMOS performance is crucial for designing
the instrument calibration program following its re-commissioning in
order to enable optimum NICMOS science.

The NICMOS warm-up monitoring program consisted of three elements 
which are summarized in Table~\ref{tab:program}:
\ben
\item{Lamp flats were taken four times a day 
in a number of filters in all three cameras. The goal was to follow 
variations of the detective quantum efficiency (DQE) 
as a function of temperature and wavelength. Because of safety concerns,
filter wheel motions were suspended when the temperature sensors on
the detector mounting cups reached $78.1\K$. At this temperature,
the sensor's analog-to-digital converters (ADC) reached their
dynamic range limit, so that accurate temperature monitoring
was no longer possible. At this point, the
BLANK filter was inserted into the beam of all three cameras, 
and only dark current measurements were performed.}
\item{Dark current exposures were taken in all three cameras during every 
orbit that was not otherwise used. This program
was designed to allow monitoring of both the dark current of the 
detectors and possible temperature-induced electronic effects, such 
as bias drifts. The dark current monitoring remained active 
until NICMOS data taking was suspended on Jan. 11, 1999 at which point 
the detectors had a temperature of about $115\K$.} 
\item{To check for possible focus variations due to changing mechanical 
stresses in the NICMOS dewar and fore optics, a star cluster was observed
twice a week until the filter wheels were fixed in the BLANK position. 
This part of the program mainly addresses changes in the 
HST/NICMOS optics, and is not relevant to detector characterization.
It is therefore not further discussed in this paper.}
\een
Except for a brief suspension of the NICMOS instrument 
on Dec. 24, 1998 for reasons unrelated to 
the warm-up, all programs executed nominally. The data, which
are available from the HST archive, were analyzed in 
near real-time fashion, and correlated with continuous 
temperature readings from various sensors in the instrument. 
Details of the
data analysis and preliminary findings are discussed in \cite{boe99}.

In this paper, we concentrate on those results that relate to the 
properties of the NICMOS-3 detector arrays. In \S~\ref{sec:tprofile}, we
briefly describe the methods for monitoring the instrument warm-up
temperature profile. The data format and the results of the 
analysis are summarized in \S~\ref{sec:findings}. 
We show the dependence of DQE, detector bias, 
saturation levels, and dark current as a function of temperature. 
While the first three items showed a behavior consistent with
expectations, the dark current exhibited an unexpected increase and
subsequent decline between $77\K$ and $85\K$, which is discussed in
\S~\ref{sec:disc}. This increase was
large enough to compromise NICMOS sensitivity, if it had to be operated
under these conditions. As a consequence, the  NASA Independent Science 
Review Committee, which met in March 1999 at STScI, recommended
suitable laboratory experiments to study further the cause of the anomaly,
to decide whether it is likely to affect NICMOS operations under the NCS,
and to investigate the possible impact on NICMOS capabilities\footnote{The
final report of the committee is available online at 
http://www.stsci.edu/observing/nicmos\_cryocooler\_isr1999.html}.
Following this recommendation, STScI, in collaboration with the NICMOS 
Instrument Definition Team at Steward Observatory, University of 
Arizona, designed a laboratory test program to investigate one
proposed explanation for the dark current anomaly. 
We describe the design of the experiment as well as its hardware, 
setup, and results in \S~\ref{sec:laboratory}. 
\section{Temperature profile of the warm-up} \label{sec:tprofile}
On Jan. 4th, 1999, NICMOS telemetry data showed a sharp rise in the
temperature readings of all three detector mounting cup sensors 
as well as all other sensors distributed across the instrument.
This event marked the complete exhaustion of the solid nitrogen
coolant after which the NICMOS instrument was no longer
in thermal equilibrium. The exact determination of the detector
temperatures as a function of time warrants some further explanation.

All three NICMOS detectors have an on-chip temperature
sensor. However, the use of these during normal NICMOS operations
produces undesirable excess noise in the data. In addition, their accuracy
is rather poor ($\approx 2\K$). Therefore, we did not make use of
the on-chip temperature sensors until all data taking was suspended. 
The overall detector temperature profile was thus 
derived from a combination of three data sources:
\bit
\item{Temperature sensors attached to the detector mounting cups for
	temperatures up to $78.1\K$ when the ADC limit was reached.}
\item{Detector bias readings from $78.1\K$ until NICMOS data taking 
	was suspended.}
\item{The on-chip sensors after the monitoring program ended.} 
\eit

Figure~\ref{f:tprofile} shows the temperature profile of the detectors 
in cameras NIC1 and NIC3. Both detectors - as well as the one in camera NIC2 
which is not shown - experienced a very similar temperature
profile. However, small differences exist, as is evident from 
the plot of the warm-up rates of NIC1 and NIC3, 
i.e. the gradient of the temperature profile in Figure~\ref{f:tprofile}. 
In particular, there is a delay of about 2 hours between the profiles
NIC1 (and NIC2) and that of NIC3. This delay can be explained by 
their different locations inside the NICMOS dewar:
the NIC3 detector is mounted at the
front end of the dewar, further away from the
N$_2$ ice, and therefore reacted earlier to the ice depletion than
the other two cameras. 

Except for this time delay, however, the rates of all 
sensors show a basically identical 
behavior: a steep increase to $\approx 7\KPD$ on Jan. 4 immediately following 
the ice depletion, a slowing of the warm-up to a rate of $\approx 5\KPD$ 
on Jan. 6, when a second rate increase to $\approx 10\KPD$ occurred. 
The most likely explanation for this second rate increase 
is outgassing from a charcoal getter inside the NICMOS dewar. 
The purpose of this getter was to adsorb any gas 
that might have permeated the NICMOS vacuum seals during the pre-launch 
cold period. Because the adsorption capacity of a charcoal getter is a strong 
function of temperature (and pressure), the getter is expected to have 
released nitrogen and oxygen throughout the NICMOS warm-up. The gas 
constitutes an additional heat conduction path towards the detectors, hence
the increase in the warm-up rate.
\section{Results of the monitoring program} \label{sec:findings}
\subsection{Data format}
All monitoring data were obtained with the 
MULTIACCUM readout sequences as described in the NICMOS
data handbook \citep{boe00}. Briefly, the MULTIACCUM scheme
is a series of non-destructive detector reads which optimize 
the dynamic range and allow accurate removal of cosmic rays throughout
the total exposure time. In addition, variations of the pixel reset
levels (or bias) can be corrected by subtracting from all further
reads a frame which is taken immediately after the detector reset. 
This frame is called the ``0th read'' in the usual NICMOS terminology.
The time intervals - often referred to as $\Delta$-times - 
between the successive reads of a MULTIACCUM sequence
can be varied. The NICMOS flight software allows a 
number of pre-programmed MULTIACCUM sequences.
A particularly important sequence which was used extensively throughout
the dark current montoring program is called STEP64. It consists of
a number of reads with logarithmically increasing $\Delta$-times up to
an integration time of $64\s$, and equally spaced $\Delta$-times of
$64\s$ beyond that.
%
\subsection{Bias levels}\label{subsec:bias}
Throughout the warm-up, the bias levels were monitored to prevent the 
signal in the high-responsivity pixels from reaching the maximum of the 
dynamic range of the ADCs. A procedure was put in place to 
quickly adjust the bias offsets, in order to prevent a significant fraction 
of pixels in the flat field data from reaching the ADC limit of 32,768 counts. 
As Figure~\ref{f:bias} shows, the mean signal in the 0th read for all three
detectors changed at a rate of $\approx \rm 280\,counts\,K^{-1}$, 
for a total change of about 15000 counts between 62 and $118\K$. 
This change did not 
necessitate any bias adjustments during the warm-up. 

Pre-launch ground testing has shown the accurate linear relationship 
between bias level and detector temperature to hold to at least $120\K$. 
This justifies the use of the bias level as a thermometer to determine the 
detector temperature in the range where no direct readings of the 
mounting cup sensors are available, as described in \S~\ref{sec:tprofile}.
\subsection{Amplifier glow}\label{subs:ampglow}
Amplifier glow is a well-known feature of NICMOS-3 arrays.
It manifests itself as a spatially variable, but highly repeatable 
signal component in every detector read-out. The signal is highest in
the corners of the array, i.e. closest to the read-out amplifiers, and
gets fainter towards the center of the array.
Typical values for the amplifier glow are 2 DN/read in the center of the array, 
and up to 15 DN/read in the corners. The signal is extremely repeatable
and can be well modeled and removed during pipeline calibration.

The amplifier glow is measured by subtracting the first two 
reads in a STEP64 sequence which are only $0.3\s$ apart. The 
subtraction eliminates any contribution from the shading profile 
(see \S~\ref{subsec:shading}), and the short integration time does not allow 
a significant signal from the linear dark current. The amplifier
glow has been measured from the data of the dark current program
throughout the instrument warm-up. In agreement with expectations,
it is constant over the entire range of our measurements. 
\subsection{Shading profile} \label{subsec:shading}
The shading profile is caused by bias variations of the read-out 
amplifiers throughout the sequential addressing of all pixels
in a detector quadrant. These bias variations have been found 
to be well-correlated with the time 
interval between readouts ($\Delta$-time), over the full temperature range 
of the warm-up. The top panel of Figure~\ref{f:shading} 
shows the variations of the shading profile in camera NIC2 with 
temperature throughout the warm-up. The median shading signal in 
columns 145 to 155 is plotted in the bottom panel of Figure~\ref{f:shading}
as a function of detector temperature. It can be well modeled by a 
second order polynomial (dash-dotted line), a fact which will
be used to remove the shading during pipeline calibration 
\citep{mon99}. Since it is a noiseless contribution to the image, it 
can be completely removed by subtracting two reads with identical 
$\Delta$-times. Subtracting the first $64\s$ $\Delta$-time read from 
the last (after removing the accumulated amplifier glow, 
as described in \S~\ref{subs:ampglow}) therefore leaves only the signal 
component which is linearly accumulated during the $1000\s$ time 
interval between those two reads. This component, which is 
discussed in \S~\ref{subsec:lineardark}, is usually referred to as the 
``true'' or ``linear'' dark current in NICMOS data.
\subsection{Detective quantum efficiency}\label{subsec:dqe}
The DQE changes as a function of temperature. 
The flat monitoring program was designed to determine the 
DQE of the NICMOS detectors at the NCS operating temperatures. 
Expectations were that most pixels would experience a 
significant increase in DQE, especially at shorter wavelengths. For 
the analysis, we used the data as processed by the CALNICA pipeline 
(``\_cal'' files). To first order, this eliminates any effects 
caused by saturation, cosmic rays, and non-linearity. The 
temperature-dependent dark current and possible sky signal do not 
affect the analysis because each dataset consists of a pair of 
``lamp off'' and ``lamp on'' exposures. Both are exposures of the (random) 
sky through a particular filter, but one has the additional signal 
from the flat field calibration lamp, which is located at the 
back of the Field Offset Mirror. Differencing these two exposures 
then leaves the true flat-field 
response from which the DQE increase relative to pre-warm-up can be 
derived. An additional complication is the fact that the pixel saturation 
levels also vary with temperature, as discussed in \S~\ref{subsec:satlevels}. 
All pixels that showed signs of saturation during the MULTIACCUM 
sequence were excluded from the analysis.

%
For all pixels, the DQE increases roughly linearly between $63\K$ 
and $78\K$, with a usually small curvature term. In all cameras, the 
linear slope is higher than average for the low-sensitivity regions, 
and lower than average for the high sensitivity regions. This behavior 
effectively flattens out the DQE variations across the array, as can 
be seen in Figure~\ref{f:ffields} which compares flat field exposures of 
all cameras through the F110W filter at 62 and $78\K$. 
The histograms on the right hand side in Figure~\ref{f:ffields} 
clearly show a smaller spread in pixel values at the higher temperature. 

Figure~\ref{f:dqevslambdatemp} summarizes the average DQE changes for all three 
NICMOS detectors as a function of both wavelength and detector temperature. 
The DQE increase is a well-behaved function of both wavelength and temperature. 
Based on the data for camera NIC3, we have linearly interpolated the measured 
DQE increase between the wavelengths of the employed filters for a number of 
temperatures. The DQE improvements are very regular
and predictable. At $75\K$, the expected NICMOS operating temperature 
under the NCS, the average responsivity increased 
by about 45\% in J-, 33\% in H-, and 17\% in K-band. 

Because of the regular behavior of the DQE, it is possible to create synthesized
flat fields at arbitrary temperatures and wavelengths by interpolating
the model fits for each detector pixel over both parameters. 
These synthetic flat fields have been extensively tested, and proven
to reliably reproduce the DQE for the temperature range of interest. 
Because the routine pipeline calibration of NICMOS data at this time 
cannot take into account temperature changes of
the detectors, the NICMOS group provides a web-based tool to
create synthetic flat field exposures for all instrument filters at
arbitrary temperatures\footnote{NICMOS users who wish to improve on 
the pipeline calibration of their data can access the tool from
the NICMOS website under http://www.stsci.edu/cgi-bin/nicmos.}.

\subsection{Saturation levels and dynamic range}\label{subsec:satlevels}
The saturation level of a given detector pixel is defined by amount of charge 
``loaded'' onto it during the detector reset. Since the reset voltage of
the NICMOS detectors is sensitive to temperature changes, the pixel saturation 
levels are expected to be a function of temperature. 
The flat field exposures taken during the course of the monitoring program
allow us to measure this effect. As summarized in Figure~\ref{f:sat} for the
NIC2 camera, the average pixel saturates earlier at higher temperatures. 
However, the intrinsic capacitance of the detector pixels is not expected to change
over the temperature range discussed here ($60\K - 80\K$). Therefore, the
resulting loss of dynamic range at higher temperatures can be compensated by 
an adjustment of the reset voltage such that full use of the pixel capacitance 
is ensured.
\subsection{Readout noise}
Each NICMOS detector has four independent readout amplifiers, each of
which reads a $128\times 128$ pixel quadrant. The noise associated
with the amplification process, commonly referred to as read noise, is
not expected to be a strong function of temperature. The dark current
monitoring data allow us to test this expectation. Subtracting the first
two reads of a STEP64 sequence eliminates all effects 
of bias variations or shading. The effective integration time
of this difference image is only $0.3\s$, too short for the 
linear dark current signal to become important. Therefore, the RMS 
deviation of the pixel values across the detector array is an accurate 
representation of the intrinsic read noise of the detectors. 
We plot the resulting read noise measurements as a function
of temperature in Figure~\ref{f:readnoise}. From these measurements, we
can confirm that the read noise is indeed fairly constant over the full
temperature range covered by our data. 

We converted the read noise from DN to electrons by using the
following conversion gains: $5.4\>{\rm e^-/DN}$ for NIC1 and NIC2, and
$6.5\>{\rm e^-/DN}$. The fact that Figure~\ref{f:readnoise} shows
camera NIC3 to have a
slightly higher readnoise (in e$^-$) than the other two cameras likely
indicates that its true conversion gain is somewhat lower than assumed. 
In order to match the read noise levels of the other two cameras, 
the NIC3 gain would have to be $5.9\>{\rm e^-/DN}$.
\subsection{Detector cosmetics}
Throughout the warm-up, no evidence was found for any significant 
changes in the detector cosmetics, i.e. the number of hot/dead pixels 
remained constant, the position and amount of grot did not change, 
and no debonding or other mechanical pixel defects were observed. 
\subsection{Linear dark current}\label{subsec:lineardark}
The linear dark current is measured after subtraction of 
amplifier glow and correction for shading. Special care was taken 
to minimize the impact of those measurements that were affected by 
high cosmic ray persistence 
after an HST passage through the South-Atlantic Anomaly. 
Figure~\ref{f:darkimages} shows 
some example exposures that demonstrate the varying structure of 
the dark current throughout the warm-up. All images are shown with 
an identical color stretch. 

The median signal of all three NICMOS arrays for the whole temperature 
range of the warm-up is plotted in Figure~\ref{f:darkprofile} on both
a linear and logarithmic scale. The notable increase 
and subsequent decline of the dark current between 77 and $85\K$ is an 
unexpected feature to which we refer as the ``bump'' for the remainder 
of this report. An increasing number of pixels with above average dark 
current is responsible for the ``salt and pepper'' appearance of the 
images in Figure~\ref{f:darkimages}.
Compared to NIC1 and NIC2, the NIC3 detector shows a much larger number 
of such ``hot'' pixels at temperatures above $85\K$. This explains the 
elevated median dark current of NIC3 compared to the other two detectors 
at temperatures above $85\K$ (Figure~\ref{f:darkprofile}).
\subsubsection{Absence of the grot}
An important observational fact is that the dark current images 
taken over the duration of the bump do not show any signs of ``grot''. 
Grot is the commonly used term for a number of small flecks of black 
paint on the detector surfaces. These particles presumably were scraped 
off the baffles during mechanical contact with the filter wheel housing, 
the process that led to the shortened NICMOS lifetime. Because grot prevents 
incoming photons from reaching the detector material, it is clearly 
visible in all NICMOS flat field images as clusters of cold pixels,
i.e. pixels with very low responsitivity. 
The dark current images obtained throughout the duration of the 
bump do not show any sign of the grot. This indicates 
that if NIR photons are indeed responsible for the bump, they must 
have come from within or behind the detector. Also, a transiently hot part
inside the NICMOS dewar is inconceivable, because all temperature 
sensors showed a monotonic increase during the warm-up. In addition, 
in order to produce such a close match to flat field exposures, the 
signal must have been produced in or close to a pupil plane, which 
does not exist between the filter wheel and the detectors. 
One can therefore conclude that the bump 
signal can not be produced by NIR photons from outside the detector.
\subsubsection{Morphology of the dark current bump}
Another important observational result is that the morphology of 
the excess signal that constitutes the dark current bump closely resembles 
the spatial variations of the DQE. This can be
most easily seen when comparing the structure in the
dark current images at the peak of the bump in Figure~\ref{f:darkimages}
to the flat field exposures shown in Figure~\ref{f:ffields}. This
similarity between dark current and DQE is only seen over the
temperature range of the bump signal. Any valid explanation for the 
dark current bump must account for this correlation, which we investigate 
in more detail in \S~\ref{subsec:corr}.

One conclusion that can be drawn from the DQE-like bump morphology is that 
the electrons responsible for the bump signal are subject to the same spatial 
variations in material properties as ``normal'' signal electrons. 
Quantities such as impurity density, mean free path length, or recombination 
efficiency all affect the detection probability of a charge carrier. 
Therefore, the bump electrons are likely to originate 
at the same physical location as signal electrons produced by infrared
photons. 

We emphasize that a temporary rise in detector temperature can be ruled out 
as the source of the enhanced dark current. This is because the signal 
morphology at dark current levels comparable to the bump 
(between 90 and $96\K$, Figure~\ref{f:darkimages}) 
is very different, and certainly does not reflect the DQE structure.
Moreover, as described in \S\ref{subsec:bias}, the detector bias levels
are very sensitive to temperature changes, but certainly show no
evidence for a transient heating of the detectors
in excess of the overall instrument warm-up (see Fig.~\ref{f:bias}).
\section{Discussion of the on-orbit dark current} \label{sec:disc}
The data described in the previous sections provide a comprehensive
study of the performance of NICMOS-3 detectors as a function of their
operating temperature. The NICMOS warm-up program offers a unique 
opportunity to study the effects of the space environment on HgCdTe detectors.
A significant unexpected result was the elevated dark current level at
temperatures between 77 and $85\K$. Because the bump is located at 
or close to the expected operating temperature for operations under 
the NCS, its nature needs to be understood for successful instrument 
calibration. In particular, it is important to determine
whether the enhanced dark current will be observed during science 
operation in Cycle 11 and beyond. Long exposures in narrow-band filters
at wavelengths below $1.7\mum$ are of particular concern, since 
for these, the NICMOS sensitivity is limited by the noise 
associated with the dark current signal. The dark
current bump observed in NICMOS therefore warrants further
investigation. 

The theoretical expectation for the dark current at temperatures 
above $140\K$ is to follow the charge carrier concentration 
\citep{coo93}, which, in turn, rises with temperature 
according to the Boltzmann factor $\rm e^{-E/kT}$. At temperatures 
between 90 and $140\K$, a generation-recombination model described 
by \cite{rog88} provides the best agreement with 
the laboratory measurements of Cooper et al. (1993). The two 
regimes both produce a basically linear relation of 
log(dark current) vs 1/T, but with different slopes. 
At temperatures below $90\K$, poorly understood tunneling effects 
are known to cause a deviation from the generation-recombination 
model. Tunneling effects cause a flattening of the dark current curve 
at colder temperatures, eventually approaching an asymptotic dark current 
level. This behavior is also evident from Figure~\ref{f:darkprofile}. 
However, none of the models predict an increase and subsequent
decline in the temperature range between 77 and $85\K$. 

As described in \S~\ref{subsec:lineardark}, 
the warm-up observations rule out both a temporary increase
in detector temperature as well as a radiative signal from outside
the detector as the cause for the bump. 
One possible origin for the charge released over the duration
of the bump is a population of electrons which
was ``trapped'' inside the detector 
material as long as it was colder than about $75\K$. 
As the detectors warmed up above this threshold, 
the trapped charge was gradually released over the
temperature range of the bump, until at about $85\K$ all traps were
emptied. The additional charge diffuses to the pn-junctions, thus giving 
rise to the enhanced signal that constitutes the bump. 
However, this qualitative scenario leaves a number of questions open,
such as the nature and number of the putative traps, the origin of
the trapped charge, and the mechanism and time constants for releasing
trapped charge.
From the evaluation of the monitoring data, one can make a few comments
that might illuminate these issues further.
\subsection{The Bump\,-\,Flat Field correlation}\label{subsec:corr}
The observed similarity of the bump morphology to 
the DQE variations suggests that the traps are distributed at a
depth inside the bulk material of the detector roughly equal
to the absorption length of infrared photons.
To illustrate this point, we show in Figure~\ref{f:det} 
a cross section of the NICMOS-3 detector.
Before infrared photons enter the active detector material, they
pass through a transparent sapphire substrate. In order to improve
the lattice match between the sapphire and the HgCdTe material, a narrow
layer of CdTe is grown between the sapphire and the HgCdTe bulk material.

Although the insertion of the CdTe layer improves the lattice match
considerably, it is far from perfect, and the CdTe-HgCdTe boundary is 
expected to contain a large number of interface traps. 
If trapped electrons were indeed
released at the CdTe-HgCdTe boundary, a general match
between the morphologies of the bump and the DQE would
be explained naturally. This is because the electrons on their
diffusion path towards the pn-junction are subject to 
variations in the carrier lifetime inside the HgCdTe layer which 
give rise to non-uniformities in the DQE across the detector.

A more detailed test can be conducted by comparing
the bump morphology to flat field exposures taken at various wavelengths.
To first order, one would predict that flat fields taken at shorter
wavelengths produce a better match to the bump structure, because
shorter wavelength photons do not penetrate as deeply into the detector
material as longer wavelength photons. They are absorbed closer to
the CdTe-HgCdTe interface, the suggested location of the traps.
If, on the other hand, the traps were distributed uniformly over the 
detector material, a flat field taken over the full responsivity range
of the detector (i.e. $0.8 - 2.5\mum$) should provide the best match 
to the bump morphology.

Discriminating between these two predictions requires flat field
exposures taken over a broad range of wavelengths {\it at the temperature
of the bump peak} at approximately $82\K$. Unfortunately, the flat
field monitoring program obtained data in only a few broad-band filters
(see Table~\ref{tab:program}) at only a number of temperatures below
$78\K$. We have therefore used the well-behaved DQE dependency on temperature 
and wavelength described in \S~\ref{subsec:dqe},
to built a set of synthetic monochromatic flat 
fields for a temperature of $82\K$, covering the full sensitivity
range of the NICMOS detectors. In Figure~\ref{f:flatmatchimages}, 
we compare the morphology of these to that of the bump. More specifically, 
we show images of the ratio between the
normalized bump signal (after subtraction of the linear dark current) and
the synthetic flat field exposures, sorted by wavelength. A spatially 
uniform ratio image means a good match between the bump signal and the 
DQE morphology. Obviously, the agreement is better for shorter wavelengths.

This result is quantified better in Figure~\ref{f:flatmatchplot} which
plots the standard deviation in the (ring-median filtered)
ratio images as a function of their wavelength. 
The ring-median filtering eliminates pixel-to-pixel variations and
emphasizes the large-scale structures in the ratio images. 
For the NIC2 and NIC3 detectors, the DQE structure and the
bump signal match best at the shortest wavelengths.
In NIC1, the standard deviation at the shortest
wavelengths is dominated by a pattern of diagonal stripes in the 
ratio images. These are likely due to the illumination pattern
of the flat field lamps.

In summary, the fact that the best match between bump morphology and DQE is
obtained at the shortest wavelengths indicates that the excess charge
detected over the course of the bump originated at or close to the 
CdTe-HgCdTe interface. If the excess charge is indeed due to a population
of traps introduced during the manufacturing process, a similar behavior 
would be expected in all NICMOS-3 detectors, at least in those from the 
same lot. The laboratory test program recommended by the Independent Science
Review Committee was designed to address this question.
In what follows, we describe the motivation, design, and
results of this test program.
\section{The ``Bump Test'' - A Laboratory Experiment}\label{sec:laboratory}
One proposed scenario for filling the traps is via normal signal electrons 
produced by infrared radiation. To test this specific hypothesis, a 
controlled experiment was conducted at the NICMOS detector laboratory at 
Steward Observatory, University of Arizona. The goal of this program
was to replicate the on-orbit warm-up profile described in 
\S~\ref{sec:tprofile}, and to measure the dark current of a NICMOS-3 
detector as a function of temperature for two scenarios. In the first,
the device was cooled down, and {\it not exposed to any external illumination
in the cold state}. Since no signal electrons were produced that could 
fill the putative traps, the expectation is that the subsequent warm-up
should not show the dark current bump. In the second scenario,
the detector was {\it flood-illuminated in the cold state}, with 
levels exceeding the charge amount under the bump by about three 
orders of magnitude. In this case, one expects the putative traps to be 
filled before the warm-up starts, and hence the bump should be reproduced.
 
Details about the major elements of the laboratory 
equipment and the test procedures may be found in \cite{boe00b}. 
Here, we only give a brief summary to illuminate some crucial aspects.
The test detector - a NICMOS-3 flight 
spare array with characteristics that are very similar to 
the on-orbit detectors - was secured in a flight-like mount and installed 
in a dewar with a cold-shuttered optical window. The dewar contains two 
filter wheels which can be externally commanded 
and rotated for optical stimulation at selected wavelengths.
A temperature-controlled stage accurately holds the array to any
desired temperature.   
The thermal background inside the dewar produces a detector signal
of about $0.3\eps$. This sets a lower limit to any dark current 
measurements, but is smaller than the actual detector dark current 
over most of the temperature range of interest.

In order to establish a baseline measurement
of the dark current increase with temperature, the detector 
was cooled to $63.4\K$ {\it without any prior illumination}
and allowed to thermally stabilize.  
After a series of dark exposures taken at this baseline temperature,
the detector was warmed to $88\K$ through a sequence of 14 linear ramp 
segments. The ramp slopes were chosen to closely follow the 
on-orbit warm-up of the NICMOS flight detectors, and replicated 
the on-orbit warm-up profile with high accuracy.
During the entire duration of the warm-up (about 5 days), STEP64 MULTIACCUM
exposures were taken continuously, each with a total exposure time 
of $1088\s$. This data format is identical to that used during the on-orbit 
monitoring program.

Following this initial warm-up, the detector was re-cooled to $63.4\K$ and 
another set of baseline darks was obtained to provide a
consistency check with the previous data. The cold detector was
then exposed to a high level of incident NIR light. Four discrete 
passbands with a FWHM of $0.1\mum$ were used to cover the full range 
of spectral sensitivity of the array\footnote{In order to assure 
quantitative knowledge of the incident 
flux level, we chose not to perform the flooding unfiltered. Also,
using a very broad bandpass would have saturated the detector 
in the shortest possible read time.}. The primary goal of 
the flooding process was to expose the detector to broad-band
illumination levels greater than the total charge released during the 
on-orbit bump anomaly, i.e. $\sim 8\times 10^5\>{\rm e^-pixel^{-1}}$.  
However, the probability of a signal electron being ``caught'' in
one of the putative charge traps is unknown. As a compromise between 
test duration and probability of filling the traps, we adopted a
$\sim$1000 times higher integrated flood signal. 
The total accumulated signal achieved over the four passbands 
during the flood was $\rm 1.7x10^{9} {\rm e^-/pixel}$, 
as detailed in Table~\ref{tab:illum}. After the flood illumination, 
another series of baseline darks was taken. The warm-up
profile and dark current measurements were then
repeated identically as in the pre-flood phase.  
If the proposed scenario for charge traps was correct, this sequence 
should have reproduced the on-orbit dark-current bump.

We also investigated any systematic effects caused by the 
detector not being in thermal equilibrium during the
warm-up. It is known, for example, that the thermal coupling 
between the array and its temperature sensor through the detector
stage is not perfect. This will introduce some 
amount of thermal lag, so that the measured dark current is actually
attributed to a slightly wrong detector temperature. 
To address these issues, the detector was again cooled to $63.4\K$ and 
allowed to thermally equilibriate. After another series of baseline darks,
the detector was warmed through a number of thermal plateaus. 
At each of these (at temperatures of 63.4, 70, 77, 83, and $88\K$), 
the detector was held stable for 5 hours. After equilibration, ten dark 
current exposures were obtained at each plateau. 
These measurements were compared to those in the
same temperature regimes, during the continuous warm-up. From this
comparison, we obtained an empirical measure of the 
heat transfer efficiency between the detector and its temperature sensor
which is discussed in the next section.
\subsection{Data Reduction \& Analysis}\label{subsec:datared}
For all MULTIACCUM sequences obtained throughout the test program,
we measured the dark current accumulation between reads
14 and 25. The 14th read was chosen as the reference frame
because it is well beyond any device non-linearities introduced by 
the reset gradient (a.k.a.``shading'') of the device \citep{rie93}.
The signal difference between the two reads divided by the net integration
time yields the linear dark current. We measured the 
three-sigma clipped mean and median dark current in each
detector quadrant over a 70x70 pixel sub-array. 
The sub-arrays were used to avoid the bulk of the amplifier glow
and other anomalies at the quadrant boundaries. The
measured dark current was then related to temperature for both the 
pre- and post-flood test runs. 

However, there is an additional complication to the interpretation of
the measurements. The above measurement does not yield the true 
dark current, because the detector is not in thermal equilibrium. 
As the device warms up, the two reads - which are $704\s$ apart - are
taken at different temperatures. This introduces a DC bias change. The
DC bias, or zero point, can be measured from the 0th read in a 
MULTIACCUM sequence which is taken only $0.2\s$ after the reset. 
For a detector in thermal equilibrium, the zero point depends linearly 
on detector temperature. In other words, NICMOS-3 detectors make excellent 
thermometers. For the test detector, we found a DC shift rate of
$\rm \sim~190\>ADU\>K^{-1}$. The apparent signal produced by this
effect was removed via the following scheme. 

Based upon the thermal slew rate at any given time during the warm-up,
one can easily calculate the temperature difference between reads
14 and 25 of each MULTIACCUM sequence, and hence predict the
excess signal in the absence of any other effects. 
However, one cannot simply subtract this predicted excess signal
from each dark current measurement. As mentioned before, the thermal 
coupling between the detector and the sensor used to control the 
detector temperature is imperfect. Indeed,
the predicted DC drift rate underestimates the 
true value. The magnitude of this effect, however, can be 
quantified by comparing the dark current measured with the detector
in thermal equilibrium at different temperatures - as
done during the last phase of the test - with those measured with
the detector transitioning through the same temperatures.
One can then empirically find a scaling factor, which should be
proportional to the thermal impedance between the detector and its 
temperature sensor. This scaling factor is to be applied to the 
predicted DC drift contribution to the measured dark current.
For our instrumental set-up, the scale factor was found to be 
$\sim~3$, as illustrated in Figure~\ref{f:labmeas}. The solid
gray line shows the expected signal excess at each segment
of the thermal ramps after scaling. When this model of the 
signal excess is subtracted from the measured dark current
(dots), the corrected dark current (black line) exactly passes
through the measurements obtained with the
device in thermal equilibrium (squares).  

As can be seen from Figure~\ref{f:labmeas}, the measured and corrected 
dark current curves for the post-flood warm-up data show no evidence of any 
bump-like signal as seen in the on-orbit warm-up. Therefore, the proposed
explanation of photoelectrons ``stored'' in charge traps that are
intrinsic to the manufacturing process of the NICMOS detectors seems
unlikely.
\section{Alternative explanations}
An alternative mechanism for producing charge traps - which uniquely
applies to detectors in space - is displacement damage from high-energy 
protons in the low-earth orbit (LEO) environment. The HST orbits the earth in
an altitude of about $600\km$ with an orbital period of about
$94\minute$. About half of its orbits pass through the South Atlantic 
Anomaly (SAA), a region where the van Allen radiation belts reach lower 
altitudes because of the asymmetry of the earth's magnetic field. 
The charged particle flux onto the HST instruments during passage through 
the SAA is much higher than during ``SAA-free'' parts of the orbit. The total
on-orbit radiation dosage of the NICMOS detectors is fairly uncertain. 
Our best estimate is based on data from the APEXRAD software \citep{gus97}. 
Assuming a total shielding of about $10\mm$ of Aluminum, we derive an upper
limit of $\rm 4\>J\,kg^{-1}$ over the entire NICMOS lifetime.
For comparison, the average total signal integrated over the 
temperature range of the bump was $\rm 5.8\cdot10^5\,e^-/pixel$.

Another possible explanation for the bump signal, which is
physically different from charge traps within the detector,
is photoluminescence. \cite{hun80} have shown that HgCdTe does
show luminescence at energy levels within the band gap via
band-to-band and donor-to-acceptor-transitions, as well as 
bound-exciton recombination. 
While the detailed mechanism remains to be identified, it
is not inconceivable that luminescence - induced either by
the thermal energy provided by the detector warm-up or by 
mechanical stress - can produce photons
inside the detector material which are subsequently registered as
the bump signal. Clearly, a more sophisticated test program is required to
investigate this theory further. 

Finally, it has been suggested that surface leakage associated with
a transiently sublimed layer on the detector might be a viable
explanation for the bump. In this context, the outgassing of the charcoal
getter mentioned in \S~\ref{sec:tprofile} might provide a source
for the contaminant layer. None of the above scenarios can be explored 
further with the limited laboratory experiment described in this paper.
\section{Summary} \label{sec:sum}
We have presented results of the NICMOS warm-up monitoring program.
Detector parameters such as quantum efficiency, dark current, 
bias, and saturation levels have been 
measured over a wide temperature range. The measurements
provide a unique database of the characteristics of
PACE-I HgCdTe detector arrays in the space environment. The data of
the NICMOS warm-up program are available from the HST archive. 

We have found an unexpected increase in dark current in all three 
NICMOS flight detectors in the temperature range between 
77 and $85\K$. We have discussed
qualitative scenarios for its explanation, including the possible
existence of a population of charge traps within the detector material.

We have reported on a laboratory experiment undertaken to
measure the dark current as a function of temperature 
in a detector of the same manufacture, pedigree, and operating 
characteristics as the flight arrays.
The program was specifically designed to investigate the trapped 
photoelectron hypothesis. The test results did not confirm predictions
of this hypothesis. The origin of the NICMOS dark current anomaly 
is thus likely to be unique to the space environment,
the way the NICMOS detectors are operated onboard HST, or a combination
of both.
\acknowledgments
We are grateful to personnel at the Rockwell Science Center and at
Ball Aerospace for their hospitality during valuable discussions on
PACE-I technology and the NICMOS instrument, respectively. 
In particular, we would like to thank K. Vural and W. Tennant at
Rockwell who helped to clarify a number of issues related to
the NICMOS-3 arrays. We are indebted to W. Burmester of Ball Aerospace 
who not only provided his wealth of expertise on the NICMOS instrument 
and detectors, but also free lodging and skiing lessons for some of us.
We also would like to thank M. Rieke, M. Stiavelli, and
T. Casselman for helpful conversations
on the possible origin of the bump.


\clearpage

 
\ifsubmode\else
\baselineskip=10pt
\fi
 

\clearpage
 
 
\ifsubmode\else
\baselineskip=14pt
\fi
 
 
\clearpage
 

\begin{deluxetable}{lccc} 
\tablenum{1}
\tablewidth{0pt}
\tablecaption{Details of the NICMOS warm-up monitoring program \label{tab:program}}
\tablehead{
\colhead{Prog. ID} & \colhead{Target} & \colhead{Purpose} & \colhead{Filter} 
        }         
\startdata
        &         &  	  & F110W (all)        \\   
 7961   & random  & DQE	  & F160W (all)	       \\    
	&         &  	  & F222M (NIC3)  \\ 
\hline   
 7962   & NGC3603 & Focus & F110W (NIC2)   	       \\   
        &         &  	  & F108N (NIC3)	       \\   
\hline   
 7963   & blank   & Dark  & BLANK   	       \\    
 (8093) &         &   	  & 	      	      
\enddata
\end{deluxetable}

\begin{deluxetable}{lcccc} 
\tablenum{2}
\tablewidth{0pt}
\tablecaption{Flood Illumination Levels \label{tab:illum}}
\tablehead{
\colhead{$\lambda_{c}$} & \colhead{DQE} & \colhead{Obs. Flux} & 
\colhead{Duration} & \colhead{Charge} \\
\colhead{[$\mum$]} & \colhead{[\%]} & \colhead{[$\rm ADU\>s^{-1}$]} & 
\colhead{[min]} & \colhead{[$\rm 10^8\>e^-pixel^{-1}$]}
        }         
\startdata
0.8 & 0.15 & 3409  & 55  & 1.32 \\   
1.2 & 0.4  & 6383  & 78  & 3.51 \\    
1.8 & 0.6  & 10936 & 68  & 5.23 \\ 
2.4 & 0.8  & 9189  & 109 & 7.05 \\   
\enddata
\end{deluxetable}

\clearpage 
\newcommand{\fctprofile}{Top: temperature profile of the on-orbit
	warm-up of cameras NIC1 and NIC3. The curve for NIC1 has 
	been shifted by 3 K to separate the two curves. Bottom:
	gradient of the temperature profile for the same two
	cameras. Camera 2 showed a similar behavior.
	\label{f:tprofile}}
\newcommand{\fcbias}{Temperature dependence of the detector bias 
	levels.\label{f:bias}} 
\newcommand{\fcshading}{Top: temperature dependence of the shading 
	profile in camera~2. Shown are cuts along the slow readout 
	direction for 
	all dark exposures throughout the warm-up, with the lowest curves 
	corresponding to the lowest temperatures. Bottom: median shading 
	signal over columns 145 to 155 (indicated by the grey arrow) as a 
	function of detector temperature.\label{f:shading}}
\newcommand{\fcffields}{Normalized flat field exposures of all NICMOS 
	detectors, taken through the F110W filter at temperatures of 
	$62\K$ (left) and $78\K$ (right). The color stretch is the same 
	for both temperatures in each camera. The histograms on the right 
	show the flattening of the arrays at the higher temperature 
	which can also be seen by comparing the images.\label{f:ffields}} 
\newcommand{\fcdqevslambdatemp}{Expected NICMOS DQE as a function 
	of wavelength and temperature for all three NICMOS cameras. The
	solid lines are linear fits to the data points.
	\label{f:dqevslambdatemp}} 
\newcommand{\fcsat}{Mean saturation level vs. temperature.\label{f:sat}} 
\newcommand{\fcreadnoise}{Read noise as a function of detector temperature
	for all three NICMOS cameras. \label{f:readnoise}} 
\newcommand{\fcdarkimages}{``Snapshot'' dark exposures of all three 
	cameras at temperatures of 68, 82, 88, and $96\K$. Note the 
	flat field morphology in all cameras at $82\K$.\label{f:darkimages}} 
\newcommand{\fcdarkprofile}{Top: median dark current signal vs. temperature 
	for all three NICMOS cameras. Bottom: Same, but plotted on a 
	logarithmic scale versus 1/T. \label{f:darkprofile} } 
\newcommand{\fcdet}{Cross section of a PACE-I detector array 
	(not to scale). \label{f:det}} 
\newcommand{\fcmatchim}{Ratio images of normalized
	bump signal and flat field response for various
	wavelengths in NIC1 (left column), NIC2 (center column), and
	NIC3 (right column). The color stretch is the same in
	all images. \label{f:flatmatchimages}} 
\newcommand{\fcmatchpl}{Standard deviation in the ring-median
	filtered ratio images of Figure~\ref{f:flatmatchimages}. The
	filtering emphasizes the large-scale variations in the images.
	 \label{f:flatmatchplot}} 
\newcommand{\fclabmeas}{Post-flood dark currents.  
	Dark currents measured during the warm-up thermal slews (points), 
	with the detector held at constant temperatures (squares), and 
	corrected (solid black line) for temperature lag
	(solid gray line) as discussed in the text.\label{f:labmeas}} 
%
 
\ifsubmode
\figcaption{\fctprofile}
\figcaption{\fcbias}
\figcaption{\fcshading}
\figcaption{\fcffields}
\figcaption{\fcdqevslambdatemp}
\figcaption{\fcsat}
\figcaption{\fcreadnoise}

\newpage
\figcaption{\fcdarkimages}
\figcaption{\fcdarkprofile}
\figcaption{\fcdet}
\figcaption{\fcmatchim}
\figcaption{\fcmatchpl}
\figcaption{\fclabmeas}
\clearpage
\else\printfigtrue\fi

\ifprintfig
\clearpage
\begin{figure}
\epsfxsize=16.0truecm
\centerline{\epsfbox{fg1a.ps}}
\centerline{\epsfbox{fg1b.ps}}
\ifsubmode
\vskip1.5truecm
\setcounter{figure}{0}
\addtocounter{figure}{1}
\centerline{Figure~\thefigure}
\else\vskip-0.3truecm\figcaption{\fctprofile}\fi
\end{figure}

\clearpage
\begin{figure}
\epsfxsize=16.0truecm
\centerline{\epsfbox{fg2.ps}}
\ifsubmode
\vskip1.5truecm
\addtocounter{figure}{1}
\centerline{Figure~\thefigure}
\else\vskip-0.3truecm\figcaption{\fcbias}\fi
\end{figure}
 
\clearpage
\begin{figure}
\epsfysize=20.0truecm
\centerline{\epsfbox{fg3.ps}}
\ifsubmode
\vskip1.5truecm
\addtocounter{figure}{1}
\centerline{Figure~\thefigure}
\else\vskip-0.3truecm\figcaption{\fcshading}\fi
\end{figure}
 
\clearpage
\begin{figure}
\epsfxsize=16.0truecm
\centerline{\epsfbox{fg4.ps}}
\ifsubmode
\vskip1.5truecm
\addtocounter{figure}{1}
\centerline{Figure~\thefigure}
\else\vskip-0.3truecm\figcaption{\fcffields}\fi
\end{figure}
 
\clearpage
\begin{figure}
\epsfysize=20.0truecm
\centerline{\epsfbox{fg5.ps}}
\ifsubmode
\vskip1.5truecm
\addtocounter{figure}{1}
\centerline{Figure~\thefigure}
\else\vskip-0.3truecm\figcaption{\fcdqevslambdatemp}\fi
\end{figure}
 
\clearpage
\begin{figure}
\epsfxsize=16.0truecm
\centerline{\epsfbox{fg6.ps}}
\ifsubmode
\vskip1.5truecm
\addtocounter{figure}{1}
\centerline{Figure~\thefigure}
\else\vskip-0.3truecm\figcaption{\fcsat}\fi
\end{figure}
 
\clearpage
\begin{figure}
\epsfysize=20.0truecm
\centerline{\epsfbox{fg7.ps}}
\ifsubmode
\vskip1.5truecm
\addtocounter{figure}{1}
\centerline{Figure~\thefigure}
\else\vskip-0.3truecm\figcaption{\fcreadnoise}\fi
\end{figure}
 
\clearpage
\begin{figure}
\centerline{\epsfbox{fg8.ps}}
\ifsubmode
\vskip1.5truecm
\addtocounter{figure}{1}
\centerline{Figure~\thefigure}
\else\vskip-0.3truecm\figcaption{\fcdarkimages}\fi
\end{figure}
 
\clearpage
\begin{figure}
\epsfxsize=16.0truecm
\centerline{\epsfbox{fg9.ps}}
\ifsubmode
\vskip1.5truecm
\addtocounter{figure}{1}
\centerline{Figure~\thefigure}
\else\vskip-0.3truecm\figcaption{\fcdarkprofile}\fi
\end{figure}
 
\clearpage
\begin{figure}
\epsfxsize=12.0truecm
\centerline{\epsfbox{fg10.ps}}
\ifsubmode
\vskip1.5truecm
\addtocounter{figure}{1}
\centerline{Figure~\thefigure}
\else\vskip-0.3truecm\figcaption{\fcdet}\fi
\end{figure}
 
\clearpage
\begin{figure}
\epsfysize=20.0truecm
\centerline{\epsfbox{fg11.ps}}
\ifsubmode
\vskip1.5truecm
\addtocounter{figure}{1}
\centerline{Figure~\thefigure}
\else\vskip-0.3truecm\figcaption{\fcmatchim}\fi
\end{figure}

\clearpage
\begin{figure}
\epsfxsize=16.0truecm
\centerline{\epsfbox{fg12.ps}}
\ifsubmode
\vskip1.5truecm
\addtocounter{figure}{1}
\centerline{Figure~\thefigure}
\else\vskip-0.3truecm\figcaption{\fcmatchpl}\fi
\end{figure}
 
\clearpage
\begin{figure}
\epsfxsize=16.0truecm
\centerline{\epsfbox{fg13.ps}}
\ifsubmode
\vskip1.5truecm
\addtocounter{figure}{1}
\centerline{Figure~\thefigure}
\else\vskip-0.3truecm\figcaption{\fclabmeas}\fi
\end{figure}
 
\fi
 
 
\end{document}